%% file: main.tex
\documentclass{nime-alternate} 

\usepackage{anonymize} 		   

\usepackage[utf8]{inputenc}

\begin{document}


\conferenceinfo{NIME'23,}{31 May–2 June, 2023, Mexico City, Mexico.}

\title{Interactive Neural Resonators}

%
%
%
\label{key}
%

\numberofauthors{3} 
%
\author{
%
%
\alignauthor
\anonymize{Rodrigo Diaz}\\
       \affaddr{\anonymize{Centre for Digital Music}}\\
       \affaddr{\anonymize{Queen Mary University of London}}\\
       \email{\anonymize{r.diazfernandez@qmul.ac.uk}}
\alignauthor
\anonymize{Charalampos Saitis}\\
       \affaddr{\anonymize{Centre for Digital Music}}\\
       \affaddr{\anonymize{Queen Mary University of London}}\\
       \email{\anonymize{c.saitis@qmul.ac.uk}}
\alignauthor
\anonymize{Mark Sandler}\\
       \affaddr{\anonymize{Centre for Digital Music}}\\
       \affaddr{\anonymize{Queen Mary University of London}}\\
       \email{\anonymize{mark.sandler@qmul.ac.uk}}
}


\maketitle

\begin{abstract}

In this work, we propose a method for the controllable synthesis of real-time contact sounds using neural resonators. Previous works have used physically inspired statistical methods and physical modelling for object materials and excitation signals. Our method incorporates differentiable second-order resonators and estimates their coefficients using a neural network that is conditioned on physical parameters. This allows for interactive dynamic control and the generation of novel sounds in an intuitive manner. We demonstrate the practical implementation of our method and explore its potential creative applications.

\end{abstract} 




\input{00_introduction}

\input{01_method}
\input{02_experiments}
\input{03_conclusion}

\bibliographystyle{abbrv}
\bibliography{references} 

\end{document}

%% file: 00_introduction.tex
\section{Introduction}

The synthesis of impact and contact sounds for real-time interaction has been examined using diverse physically-inspired methods. Often the problem is divided into three smaller ones - modelling or analysing a resonating body, modelling the interaction with the body and modelling the object's sound propagation in space.  

The division of the action and the object in the synthesis process has a psychoacoustic justification and has been used in several works~\cite{conan_intuitive_2014, avanzini_interactive_2005}. Similarly, implementing such a paradigm is usually realised using the source-filter approach, where the source is an excitation signal, and the filter is a bank of resonators~\cite{van_den_doel_foleyautomatic_2001}. 

The resonators, in turn, are tuned to the natural modal frequencies of the object. Obtaining such frequencies can be realised by signal-analysis techniques~\cite{cook_physically_1997,aramaki_analysis-synthesis_2006}, or through modal analysis using the \textit{finite element method} (FEM)~\cite{obrien_synthesizing_2002}. While the latter method is accurate, it is computationally demanding and must be computed anew for objects of different shapes and materials. Furthermore, when the object is assumed to be non-rigid, its mode shapes change under large deformations, and it is necessary to employ non-linear alternatives~\cite{obrien_synthesizing_2001}.

Finally, to simulate the sound propagation in space, it is necessary to determine the acoustic transfer function of the object, which can be computationally intensive as it relies on the object's specific geometry~\cite{james_precomputed_2006}.

\section{Related Work}

In the context of real-time interaction, several techniques have been proposed to improve the speed and efficiency of the modes computation~\cite{bonneel_fast_2008} and synthesis, such as using statistical methods to generate realistic finite impulse responses and their interpolation for different impact positions~\cite{traer_perceptually_2019}.

More recently, neural-based methods have been used to predict modal gains based on an object's shape and certain material parameters, where the sound is rendered using an oscillator bank~\cite{jin_deep-modal_2020, jin_neuralsound_2022}. Our present work builds on a previous approach~\cite{\anoncite{diaz_rigid-body_2022}} that uses a neural network and differentiable resonators to learn the frequency response of different shapes and materials without explicitly assuming a modal model.

Similarly, different methods have been proposed for modelling the interaction with the object depending on its type. For impact sounds, a simple model is to model the contact force on the object as a raised cosine function~\cite{van_den_doel_foleyautomatic_2001}. Non-linear models that consider the local geometric deformation on the surface of the impacted object (i.e., the extended Hertzian contact model) and of continuous frictional contact have been implemented in the works of Avanzini et al.~\cite{avanzini_controlling_2001, avanzini_interactive_2005} and Papetti~\cite{papetti_numerical_2011}. For scraping and sliding sounds, it is also important to consider the micro-collisions between a textured surface and a scraper. The microscopic surface irregularities that constitute a \textit{surface texture} can be modelled using a fractal-noise~\cite{van_den_doel_foleyautomatic_2001, rocchesso_sounding_2003}. However, these textures can be obtained from experimental observations~\cite{traer_perceptually_2019, agarwal_object-based_2021}, and alternatively, it is also possible to combine both approaches~\cite{ren_synthesizing_2010}, to enhance the realism of such textures. Our method builds upon some of these techniques for synthesising sliding and scraping sounds.  

The present work has two main contributions. The first is a practical implementation of a neural-based pipeline for real-time interactive exploration of contact sounds using physical parameters. We focus mainly on impact and scraping interaction on thin membrane-like objects of arbitrary shape and material. Second, we show an analysis of the effect of the modulation of these parameters on the sound, especially regarding their creative use.

%% file: 01_method.tex
\section{Implementation}

\begin{figure*}[!ht]
     \includegraphics[trim={0.5cm 0.5cm 0.5cm 0.5cm},clip, width=0.4\textwidth]{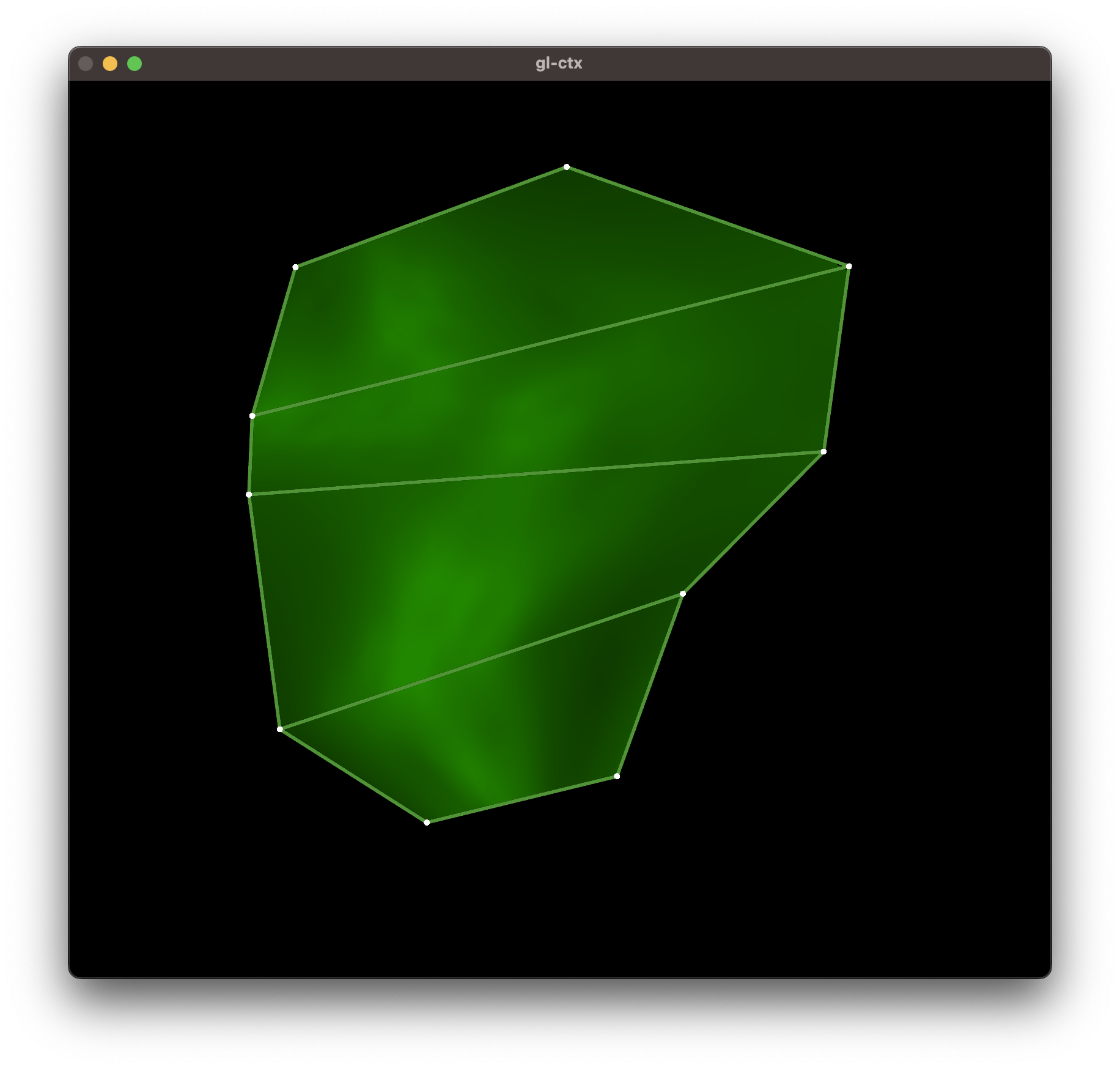}
     \hfill
     \includegraphics[trim={0.5cm 0.5cm 0.5cm 0.5cm},clip,width=0.45\textwidth]{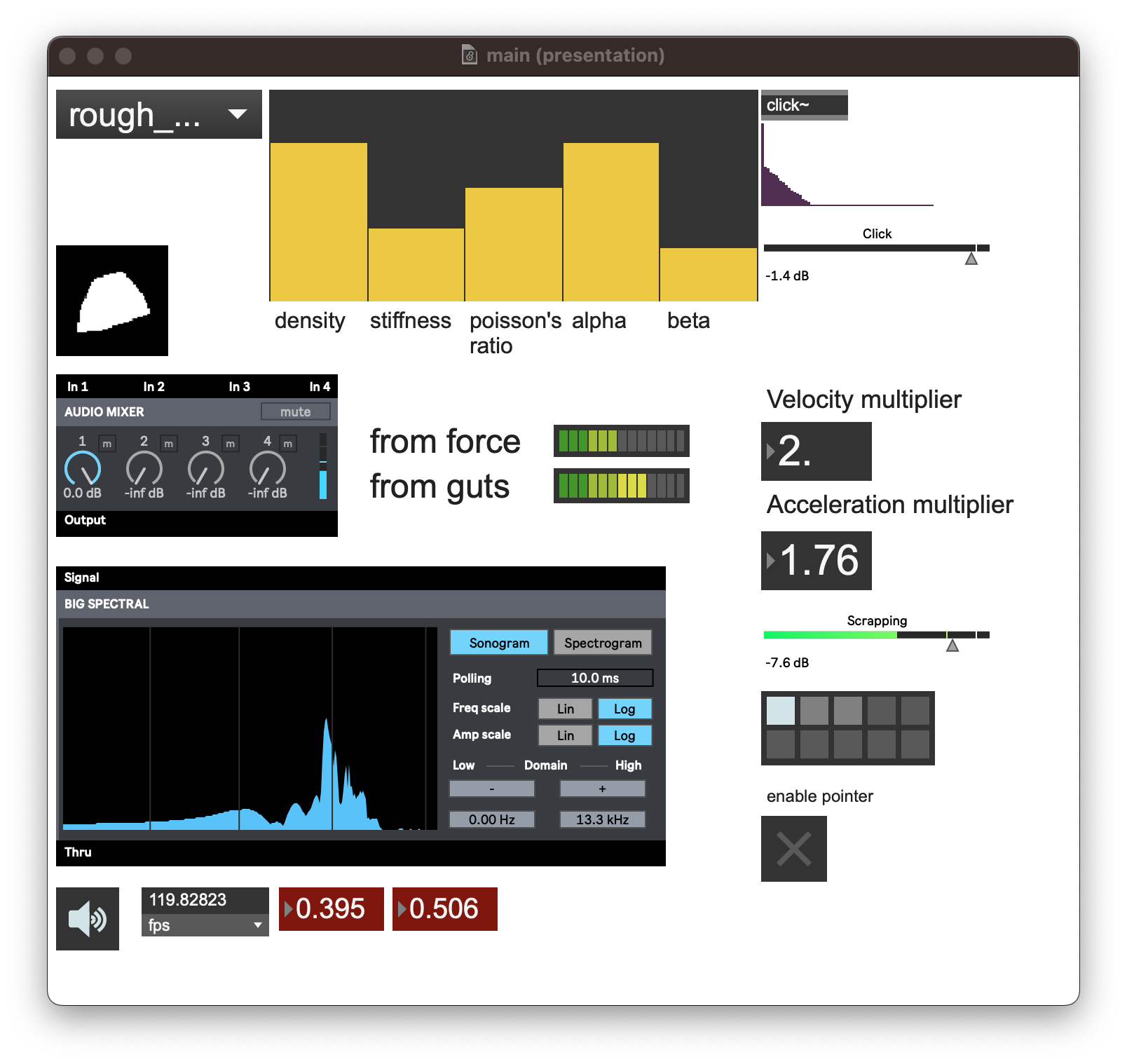}
    \caption{Prototypical user interface in Max/MSP. On the left, the interface for shape deformation is displayed. The user is able to deform a 2D mesh by moving vertex handles or generating a random mesh. They can also interact with the resonator by sliding or scraping inside the shape with the mouse or by clicking an arbitrary point (hitting). On the right, the interface for controlling excitation and material parameters is shown.}
    \label{fig:screenshot}
\end{figure*}

To model the response of a resonating body, we base our approach on a previous method~\cite{\anoncite{diaz_rigid-body_2022}}. For this, a neural network is trained using a 2D surface with an arbitrary shape and material parameters. The network predicts the filter coefficients of a differential second-order filter bank. The filter bank's frequency response is compared to a target response obtained from an object using a FEM solver. The network parameters are optimized based on the difference between the reconstructed response and the target.

For its use in real-time, the neural resonator is implemented as a collection of three \textit{externals} in Max/MSP\footnote{\url{https://cycling74.com/}}. However, since the resonator is written in C\texttt{++}, it can be easily adapted for its use in other interactive applications such as Pure Data\footnote{\url{https://puredata.info/}} or Unity 3D\footnote{\url{https://unity.com/}}. The interface for mesh deformation is implemented using OpenGL externals and Jitter. In Figure~\ref{fig:screenshot}, a screenshot of the user interface of the implementation is shown. 

\subsection{Max/MSP Externals}
\label{subsec:externals}

The first external, \texttt{encoder}, is a 2D convolutional encoder based on EfficientNet~\cite{tan_efficientnet_2019}. It takes as an input a $64 \times 64$ grid of binary values that correspond to the two-dimensional shape and outputs a vector $\mathbf{v} \in \mathbb{R}^{1000}$ that encodes the shape information. This is concatenated with two additional vectors - a position vector $\mathbf{x} \in \mathbb{R}^2$ and a material parameter vector $\mathbf{\phi} \in \mathbb{R}^5$ that comprises material density ($\rho$), Young's modulus ($E$), Poisson's ratio ($\nu$) and damping coefficients ($\alpha, \beta$). The resulting vector is normalized in the range $[0, 1]$. The ranges for the unnormalized material parameters are set in advance, and these are $\rho \in [500, 15000]$, $E \in [8\times 10^9, 5\times 10^{10}]$, $\nu \in [0.1, 0.4]$, $\alpha \in [1, 10]$ and $\beta \in [3\times 10^{-7}, 2\times 10^{-6}]$.

The second part of the pipeline is a fully-connected network (\texttt{fc}) external. It receives the concatenated vector previously computed by the \texttt{encoder} and outputs a set of $5 \times M \times L$ coefficients, where $M$ is the cascade depth of the filter (second order sections), and $L$ is the number of parallel filters. Then each section transfer function is given by:

$$
H_{l,m}(z)=g \frac{1+b_{1, l, m} z^{-1}+b_{2, l, m} z^{-2}}{1+a_{1, l, m} z^{-1}+a_{2, l, m} z^{-2}}
$$

The coefficients are fed to a configurable \mbox{\texttt{filterbank\textapprox}} external that implements the second-order resonators using the Direct-Form-II realization~\cite{smith_introduction_2007}. The modular design of the pipeline enables us to predict sounds based on changes in position and material parameters separately from changes in shape, reducing the computational load required for the predictions.

\subsection{Excitation signals}

The resonator bank can be excited in different ways. In the case of impact sounds, we use a kaiser window-like impulse with a controllable $\beta$ parameter~\cite{zambon_accurate_2012}. This single parameter can be used to control the apparent hardness of the impact. Note that methods for non-linear contact excitation~\cite{avanzini_controlling_2001} can also be applied with a modified configuration of the \mbox{\texttt{filterbank\textapprox}}~\cite{zambon_accurate_2012}. Alternatively, the user can also manually draw the shape of the impulse as shown in Figure~\ref{fig:screenshot}.

For the simulation of scraping sounds, a method based on previous works~\cite{agarwal_object-based_2021,traer_perceptually_2019} is implemented. To simulate surfaces of different roughness and structure, we utilize the \textit{pre-quilted} signals~\cite{efros_image_2001} provided by the authors\footnote{\url{https://github.com/threedworld-mit/tdw}}. Assuming the contact with surface $S(x,y)$ is continuous, we compute the vertical and horizontal forces $F_v, F_h$ given by:

$$
\begin{aligned}
F_{\mathrm{v}}(t) &= m_{\mathrm{p}}\left(\frac{\partial^2 S(x, y)}{\partial x^2}\left|v_x(t)\right|^2+\frac{\partial^2 S(x, y)}{\partial y^2}\left|v_y(t)\right|^2\right) \\
F_{\mathrm{h}}(t) &= v_x(t) \frac{\partial S(x, y)}{\partial x}+v_y(t) \frac{\partial S(x, y)}{\partial y} \\
F &= \alpha F_{\mathrm{v}} + \beta F_{\mathrm{h}} 
\end{aligned}
$$

where $v_{x}(t), v_{y}(t)$ is the velocity scraper in each direction and $m$ is the mass of the scraper. The parameters $\alpha$ and $\beta$ are set arbitrarily by the user to scale the vertical and horizontal forces, respectively.

\subsection{Control Parameters}

Inference of coefficients for different materials, positions and shapes can be computed in real-time~\cite{\anoncite{diaz_rigid-body_2022}}, allowing dynamic modulation of the parameters.

When the user creates or modifies a shape, each position within the shape results in a different set of coefficients for the \mbox{\texttt{filterbank\textapprox}} object. Positional values are in the \textit{window} coordinate space and are scaled to the range [0,1], where the origin is located at the bottom left of the window. These values are streamed at Max's \textit{control-rate} (typically 1000Hz). The network predicts smooth coefficient values (as shown in Fig.~\ref{fig:modulate_position}); thus, additional interpolation of the coefficients is not strictly necessary for a smooth transition in the output sound, unlike previous works~\cite{agarwal_object-based_2021}. At the same time, the position and velocity are also used to generate a corresponding signal to excite the newly configured \mbox{\texttt{filterbank\textapprox}}.

Similarly, we can simulate non-linear behaviour by modulating the material parameters. For example, we can modulate the Young modulus manually to roughly approximate dynamic elasticity after an impact. This could be used to approximate the \textit{pitch glide}~\cite{kirby_evolution_2021} effect observed after large impacts due to non-linear materials and large geometric deformation. Naturally, it can also be modulated in more creative and less physically plausible ways (as shown in Figure~\ref{fig:all_parameters}). 

%% file: 02_experiments.tex
\section{Experiments and results}

The implementation, further results and an interactive demo can be found at \url{https://interactive-neural-resonators.com}.

\begin{figure}[h]
    \includegraphics[
        trim={.6cm .7cm .7cm .7cm},
        clip,
        width=\columnwidth
        ]{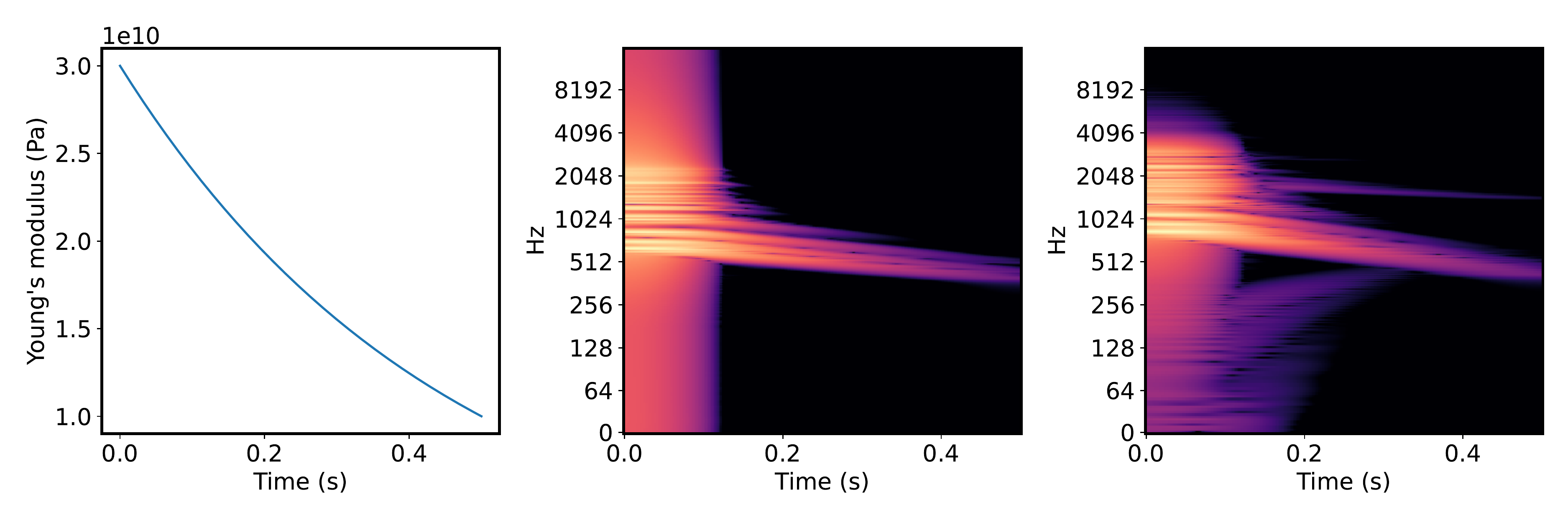}
	\caption{Dynamic control of the Young's Modulus parameter. Left, we show the parameter curve. The middle and rightmost plots show the spectrogram of the prediction and the FEM results, respectively.}
	\label{fig:young_modulus}
\end{figure}

In Figure~\ref{fig:young_modulus}, we demonstrate the effect of modulating the Young's Modulus parameter. It should be noted that the network was not trained with dynamically modulated input. The reference results, in this case, were obtained by solving a different system at each modulation step and rendering the sound using an impulse. The final modulated result is generated by combining the individual renderings using the overlap-add method. Conversely, the results obtained using our network were achieved by simply filtering an impulse and modulating the parameters at each time step.

\begin{figure}[h]
    \includegraphics[
        trim={.6cm .7cm .7cm .7cm},
        clip,
        width=\columnwidth
    ]{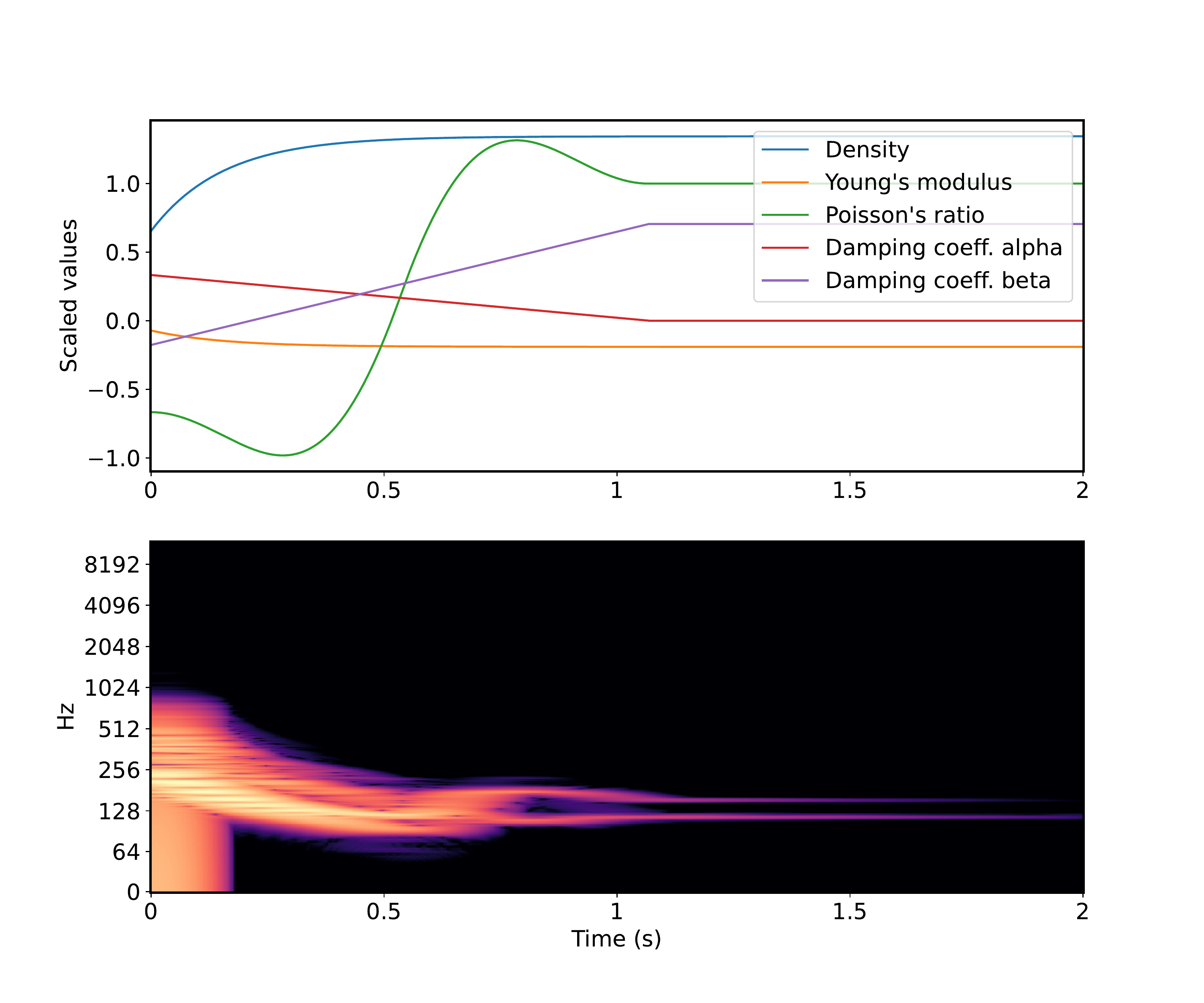}
	\caption{Simultaneous modulation of the parameters of the neural resonator model. On top, a visualization of the modulated parameters in time, where the y-axis is the scaled value of the parameters according to the ranges described in Section~\ref{subsec:externals}. The parameters are modulated beyond the [0,1] range. On the bottom a spectrogram of the generated sound.}
	\label{fig:all_parameters}
\end{figure}

The network can also extrapolate to some degree outside the training range. For example, in Figure~\ref{fig:all_parameters}, we show the simultaneous modulation of all material parameters in time. Position parameters can be modulated likewise over time. 

Modulating material and positional parameters within the training range yields results consistent with the FEM simulation. However, as the modulation falls outside this range, the predicted sound increasingly deviates from the one produced using the numerical simulation, i.e., the error with respect to the target frequency response increases. Despite these inaccuracies, the network's predictions confirm that it has acquired an understanding of certain aspects of the sounds. For example, changes in density and Young's modulus influence the pitch, while variations in Poisson's ratio affect the perceived inharmonicity. The damping parameters $\alpha$ and $\beta$ also correspond to longer decay and less damped frequencies, respectively, when outside the training range.

\begin{figure}[h]
    \includegraphics[
        trim={.6cm .7cm .7cm .7cm},
        clip,
        width=\columnwidth
    ]{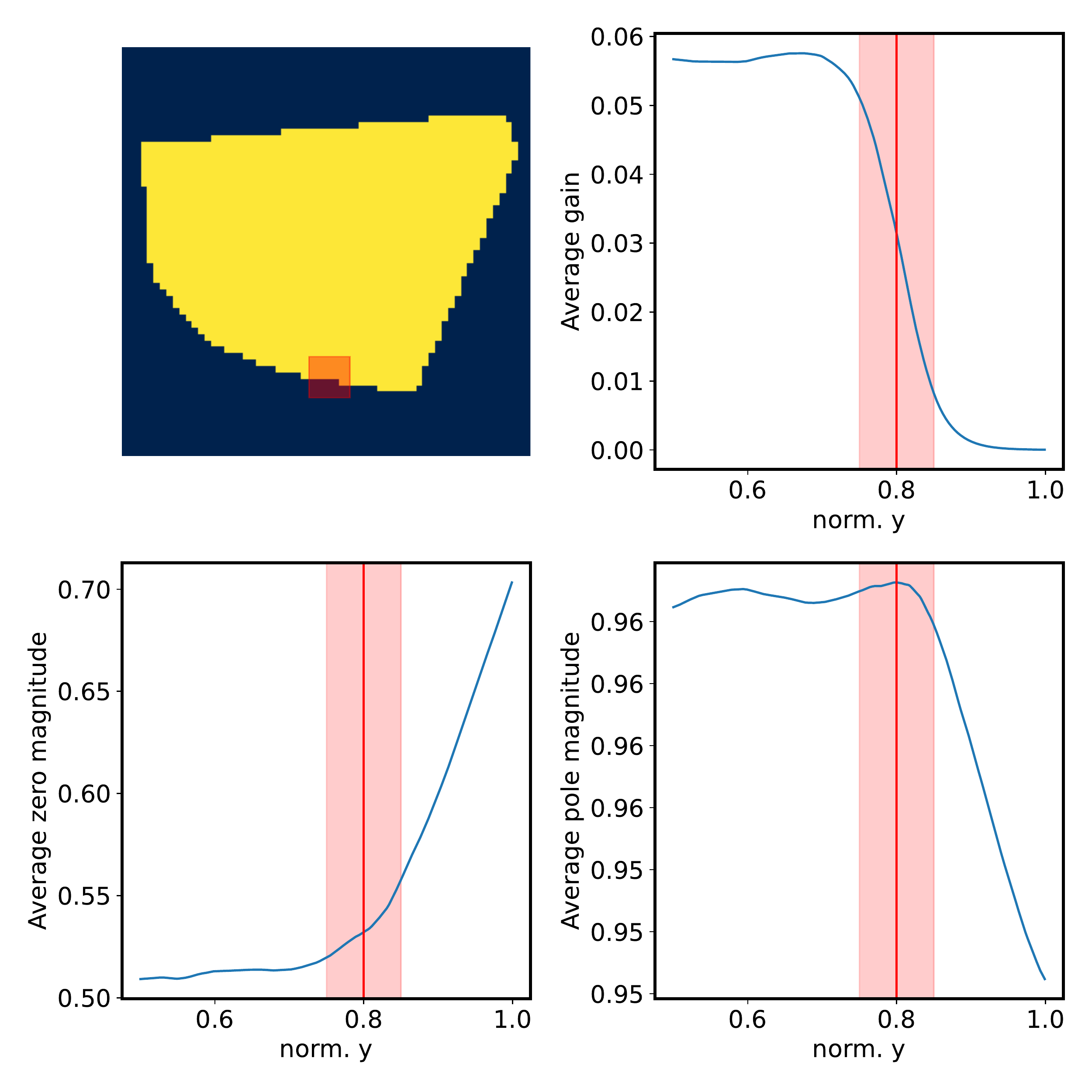}
	\caption{Average magnitude of the estimated parameters (gains, zeros and poles), as a function of the normalized vertical coordinate position. The top left panel displays the shape used for inference with the shaded region indicating the boundary's location. The estimated parameters are displayed on the top right and bottom panels, with the solid red line representing the shape's boundary and the shaded area indicating the region within the shape adjacent to the boundary (corresponding to the shaded patch in the top left panel).}
	\label{fig:modulate_position}
\end{figure}

An interesting effect is observed at the boundary and outside the shape. During training, the network is supervised using nodal positions inside the boundaries, as interaction outside the shape is physically impossible; therefore, the border itself is a discontinuity not present during training. However, the network appears to learn a \textit{softer} approximation to the function near the boundary, as shown in Figure~\ref{fig:modulate_position}. This, in turn, generates interesting ringing effects in the output sound if one scrapes rapidly outside the shape. 

%% file: 03_conclusion.tex
\section{Conclusion and Future Work}

We have presented an interactive method for synthesizing various contact sounds using neural network-controlled resonators and a practical implementation of this pipeline in Max/MSP. We believe this implementation can be used as a neural instrument for exploratory music performance and sonic art. To gain more insights into the expressive and subtle design possibilities of our method, we plan to conduct a user study where participants will be assigned \textit{tuning}, and \textit{sound-matching tasks}~\cite{armitage_subtlety_2022} using our neural instrument.

Given that the network architecture of our pre-trained model does not require significant computational resources, it can easily be adapted to run on embedded devices such as Bela\footnote{\url{https://bela.io/}}. We plan to develop an embedded version of our pipeline and complement the interaction with our neural synthesizer using audio excitation signals generated from hardware devices such as the Tickle~\cite{neupert_interacting_2019} instead of modelling physical excitation signals. This would increase the accessibility, interactivity, and frugality of our approach. Furthermore, we plan to adapt our model for inference on the web and mobile devices.

We also intend to address the shortcomings of generalization outside the domain of our training set and its robustness against approximation errors in the synthetic training set (e.g., refinement of the tessellation of the input mesh and finite element type). We believe this could be alleviated primarily through the use of real-world labelled recordings.

While we have shown the ability to control the parameters of our neural resonator manually, we believe that it could also be helpful to approximate the parameters using other target sounds as a starting point for creative exploration. This could be achieved, for example, by freezing our model's parameters and optimizing material parameters using gradient descent. Similarly, we plan to explore the possibility of optimizing the vertex parameters of an initial template mesh as a starting point for shape deformation.

\section{Acknowledgements}

\anonymize{Rodrigo Diaz is supported by UK Research and Innovation [grant number EP/S022694/1].}

\section{Ethical Standards}

The focus of this paper is the implementation of a software pipeline utilizing neural techniques trained on numerical simulations of vibration, with no ethical implications or considerations arising from the research and content presented.

%% file: main.bbl
\begin{thebibliography}{10}

\bibitem{agarwal_object-based_2021}
V.~Agarwal, M.~Cusimano, J.~Traer, and J.~McDermott.
\newblock Object-{Based} {Synthesis} of {Scraping} and {Rolling} {Sounds}
  {Based} on {Non}-{Linear} {Physical} {Constraints}.
\newblock In {\em 2021 24th {International} {Conference} on {Digital} {Audio}
  {Effects} ({DAFx})}, pages 136--143, Vienna, Austria, Sept. 2021. IEEE.

\bibitem{aramaki_analysis-synthesis_2006}
M.~Aramaki and R.~Kronland-Martinet.
\newblock Analysis-synthesis of impact sounds by real-time dynamic filtering.
\newblock {\em IEEE Transactions on Audio, Speech, and Language Processing},
  14(2):695--705, Mar. 2006.
\newblock Conference Name: IEEE Transactions on Audio, Speech, and Language
  Processing.

\bibitem{armitage_subtlety_2022}
J.~Armitage.
\newblock {\em Subtlety and detail in digital musical instrument design}.
\newblock Thesis, Queen Mary University of London, Apr. 2022.

\bibitem{avanzini_controlling_2001}
F.~Avanzini and D.~Rocchesso.
\newblock Controlling material properties in physical models of sounding
  objects.
\newblock 2001.

\bibitem{avanzini_interactive_2005}
F.~Avanzini, S.~Serafin, and D.~Rocchesso.
\newblock Interactive {Simulation} of rigid body interaction with
  friction-induced sound generation.
\newblock {\em IEEE Transactions on Speech and Audio Processing},
  13(5):1073--1081, Sept. 2005.

\bibitem{bonneel_fast_2008}
N.~Bonneel, G.~Drettakis, N.~Tsingos, I.~Viaud-Delmon, and D.~James.
\newblock Fast modal sounds with scalable frequency-domain synthesis.
\newblock {\em ACM Transactions on Graphics}, 27(3):1--9, Aug. 2008.

\bibitem{conan_intuitive_2014}
S.~Conan, E.~Thoret, M.~Aramaki, O.~Derrien, C.~Gondre, S.~Ystad, and
  R.~Kronland-Martinet.
\newblock An {Intuitive} {Synthesizer} of {Continuous}-{Interaction} {Sounds}:
  {Rubbing}, {Scratching}, and {Rolling}.
\newblock {\em Computer Music Journal}, 38(4):24--37, Dec. 2014.

\bibitem{cook_physically_1997}
P.~R. Cook.
\newblock Physically {Informed} {Sonic} {Modeling} ({PhISM}): {Synthesis} of
  {Percussive} {Sounds}.
\newblock {\em Computer Music Journal}, 21(3):38, 1997.

\bibitem{diaz_rigid-body_2022}
R.~Diaz, B.~Hayes, C.~Saitis, G.~Fazekas, and M.~Sandler.
\newblock Rigid-{Body} {Sound} {Synthesis} with {Differentiable} {Modal}
  {Resonators}, Oct. 2022.
\newblock arXiv:2210.15306 [cs, eess].

\bibitem{efros_image_2001}
A.~A. Efros and W.~T. Freeman.
\newblock Image quilting for texture synthesis and transfer.
\newblock In {\em Proceedings of the 28th annual conference on {Computer}
  graphics and interactive techniques}, pages 341--346. ACM, Aug. 2001.

\bibitem{james_precomputed_2006}
D.~L. James, J.~Barbič, and D.~K. Pai.
\newblock Precomputed acoustic transfer: output-sensitive, accurate sound
  generation for geometrically complex vibration sources.
\newblock {\em ACM Transactions on Graphics}, 25(3):987--995, July 2006.

\bibitem{jin_deep-modal_2020}
X.~Jin, S.~Li, T.~Qu, D.~Manocha, and G.~Wang.
\newblock Deep-{Modal}: {Real}-{Time} {Impact} {Sound} {Synthesis} for
  {Arbitrary} {Shapes}.
\newblock In {\em Proceedings of the 28th {ACM} {International} {Conference} on
  {Multimedia}}, pages 1171--1179, Seattle WA USA, Oct. 2020. ACM.

\bibitem{jin_neuralsound_2022}
X.~Jin, S.~Li, G.~Wang, and D.~Manocha.
\newblock {NeuralSound}: learning-based modal sound synthesis with acoustic
  transfer.
\newblock {\em ACM Transactions on Graphics}, 41(4):1--15, July 2022.

\bibitem{kirby_evolution_2021}
T.~Kirby and M.~Sandler.
\newblock The evolution of drum modes with strike intensity: {Analysis} and
  synthesis using the discrete cosine transform.
\newblock {\em The Journal of the Acoustical Society of America},
  150(1):202--214, July 2021.

\bibitem{neupert_interacting_2019}
M.~Neupert and C.~Wegener.
\newblock Interacting with digital resonators by acoustic excitation.
\newblock 2019.

\bibitem{obrien_synthesizing_2001}
J.~F. O'Brien, P.~R. Cook, and G.~Essl.
\newblock Synthesizing sounds from physically based motion.
\newblock In {\em Proceedings of the 28th annual conference on {Computer}
  graphics and interactive techniques - {SIGGRAPH} '01}, pages 529--536, Los
  Angeles, California, United States of America, 2001. ACM Press.

\bibitem{obrien_synthesizing_2002}
J.~F. O'Brien, C.~Shen, and C.~M. Gatchalian.
\newblock Synthesizing sounds from rigid-body simulations.
\newblock In {\em Proceedings of the 2002 {ACM} {SIGGRAPH}/{Eurographics}
  symposium on {Computer} animation}, {SCA} '02, pages 175--181, New York, NY,
  USA, July 2002. Association for Computing Machinery.

\bibitem{papetti_numerical_2011}
S.~Papetti, F.~Avanzini, and D.~Rocchesso.
\newblock Numerical {Methods} for a {Nonlinear} {Impact} {Model}: {A}
  {Comparative} {Study} {With} {Closed}-{Form} {Corrections}.
\newblock {\em IEEE Transactions on Audio, Speech, and Language Processing},
  19(7):2146--2158, Sept. 2011.

\bibitem{ren_synthesizing_2010}
Z.~Ren, H.~Yeh, and M.~C. Lin.
\newblock Synthesizing contact sounds between textured models.
\newblock In {\em 2010 {IEEE} {Virtual} {Reality} {Conference} ({VR})}, pages
  139--146, Boston, MA, USA, Mar. 2010. IEEE.

\bibitem{rocchesso_sounding_2003}
D.~Rocchesso, R.~Bresin, and M.~Fernstrom.
\newblock Sounding objects.
\newblock {\em IEEE Multimedia}, 10(2):42--52, Apr. 2003.

\bibitem{smith_introduction_2007}
J.~O. Smith.
\newblock {\em Introduction to {Digital} {Filters}: {With} {Audio}
  {Applications}}.
\newblock W3K Publishing, 2007.

\bibitem{tan_efficientnet_2019}
M.~Tan and Q.~Le.
\newblock {EfficientNet}: {Rethinking} {Model} {Scaling} for {Convolutional}
  {Neural} {Networks}.
\newblock In {\em Proceedings of the 36th {International} {Conference} on
  {Machine} {Learning}}, pages 6105--6114. PMLR, May 2019.
\newblock ISSN: 2640-3498.

\bibitem{traer_perceptually_2019}
J.~Traer, M.~Cusimano, and J.~H. McDermott.
\newblock A perceptually inspired generative model of rigid-body contact
  sounds.
\newblock 2019.

\bibitem{van_den_doel_foleyautomatic_2001}
K.~van~den Doel, P.~G. Kry, and D.~K. Pai.
\newblock {FoleyAutomatic}: physically-based sound effects for interactive
  simulation and animation.
\newblock In {\em Proceedings of the 28th annual conference on {Computer}
  graphics and interactive techniques - {SIGGRAPH} '01}, pages 537--544, Not
  Known, 2001. ACM Press.

\bibitem{zambon_accurate_2012}
S.~Zambon.
\newblock {\em Accurate {Sound} {Synthesis} of {3D} {Object} {Collisions} in
  {Interactive} {Virtual} {Scenarios}}.
\newblock {PhD} thesis, Universita degli Studi di Verona, 2012.

\end{thebibliography}
